\documentclass[11pt,a4paper]{article}
\usepackage[hyperref]{acl2021}
\usepackage{times}
\usepackage{graphicx}
\usepackage{amsmath}
\usepackage{amsfonts}
\usepackage{amsthm}
\usepackage{booktabs}
\usepackage{CJKutf8}
\usepackage{bm}
\usepackage{multirow}
\usepackage{tabularx}
\usepackage{enumitem}
\usepackage{float}
\usepackage{caption}
\usepackage{subcaption}

\usepackage{microtype}

\aclfinalcopy 

\newcommand{\eg}{\emph{e.g.}}
\newcommand{\ie}{\emph{i.e.}}
\newcommand{\etc}{\emph{etc}}

\title{\textit{DeepRapper}: Neural Rap Generation with Rhyme and Rhythm Modeling}

\author{Lanqing Xue\textsuperscript{\rm 1},
    Kaitao Song\textsuperscript{\rm 2},
    Duocai Wu\textsuperscript{\rm 3},
    Xu Tan\textsuperscript{\rm 4\thanks{$^*$ Corresponding author: Xu Tan, xuta@microsoft.com}}, \\
 \bf{Nevin L. Zhang\textsuperscript{\rm 1},
    Tao Qin\textsuperscript{\rm 4}, 
    Wei-Qiang Zhang\textsuperscript{\rm 5}, 
    Tie-Yan Liu\textsuperscript{\rm 4}} \\
 \textsuperscript{\rm 1}The Hong Kong University of Science and Technology\\
 \textsuperscript{\rm 2}Nanjing University of Science and Technology\\
 \textsuperscript{\rm 3}Fudan University \quad 
 \textsuperscript{\rm 4}Microsoft Research Asia \quad
 \textsuperscript{\rm 5}Tsinghua University \\
 \texttt{\{lxueaa,lzhang\}@cse.ust.hk} \\
 \texttt{kt.song@njust.edu.cn} \quad \texttt{dcwu18@fudan.edu.cn} \\
 \texttt{\{xuta, taoqin, tyliu\}@microsoft.com} \quad
 \texttt{wqzhang@tsinghua.edu.cn}
}

\date{}

\begin{document}
\maketitle
\begin{abstract}
Rap generation, which aims to produce lyrics and corresponding singing beats, needs to model both rhymes and rhythms. Previous works for rap generation focused on rhyming lyrics but ignored rhythmic beats, which are important for rap performance. In this paper, we develop \textit{DeepRapper}, a Transformer-based rap generation system that can model both rhymes and rhythms. Since there is no available rap dataset with rhythmic beats, we develop a data mining pipeline to collect a large-scale rap dataset, which includes a large number of rap songs with aligned lyrics and rhythmic beats. Second, we design a Transformer-based autoregressive language model which carefully models rhymes and rhythms. Specifically, we generate lyrics in the reverse order with rhyme representation and constraint for rhyme enhancement and insert a beat symbol into lyrics for rhythm/beat modeling. To our knowledge, DeepRapper is the first system to generate rap with both rhymes and rhythms. Both objective and subjective evaluations demonstrate that DeepRapper generates creative and high-quality raps with rhymes and rhythms. Code will be released on GitHub~\footnote{ https://github.com/microsoft/muzic/tree/main/deeprapper}.
\end{abstract}

\section{Introduction}
Rap is a musical form originating from America in 1970s, and has quickly developed as one of the mainstream music genres in the world~\cite{keyes2004rap}. With the rapid development of artificial intelligence, automatic rap lyrics generation has drawn attention from academia~\cite{potash2015ghostwriter,malmi2016dopelearning,liang2018attae,nikolov2020rapformer}. Generally speaking, rap lyrics need to be semantically meaningful and fashionable to convey interesting stories or express feelings. Different from natural language or other artistic genres (\eg, lyrics or poetry), rap has distinctive characteristics: 1) it usually contains complex rhyme patterns among several consecutive sentences, which are the key to form a good flow; 2) it needs to align with the singing beat since rap lyrics are usually rapped according to some rhythmic accompaniments. Therefore, how to generate rap lyrics with good rhymes and rhythms is a troublesome problem.

Previous works~\cite{potash2015ghostwriter,malmi2016dopelearning,liang2018attae,nikolov2020rapformer} for rap generation mainly focused on lyric generation and some of them developed strategies for rhyme modeling. \citet{potash2015ghostwriter} directly added a ``$<$endLine$>$" token at the end of verse lines and expected to learn rhyme patterns implicitly. \citet{nikolov2020rapformer} applied a two-step strategy, which first generates rap lyrics and then adds rhyme tokens to the end of generated lyrics. However, these methods cannot guarantee the rhyme patterns for every lyric line and only care the rhyme on the last token. Although many works have studied rhyming modeling in other artistic genres (\eg, poetry)~\cite{Piji2020SongNet,van2020automatic,liu2020deep}, they are not suitable for rap generation due to the complex rhyme structure in rap. For example, poetry needs to rhyme with only the last word in each sentence, while rap rhymes with multiple consecutive tokens at the end of each sentence.

No previous works have studied rhythm modeling (\ie, beats in rap), to our knowledge. One of the main reasons is the lack of rap datasets with beat-lyric alignment. Consequently, the generation of lyrics without rhythmic beats cannot be regarded as a full rap generation. 

In this paper, we develop DeepRapper, a Transformer~\cite{Vaswani2017transformer} based rap generation system which can model both rhymes and rhythms. To build the system, since there is no available rap datasets with aligned rhythmic beats, we design a data mining pipeline and collect a large-scale rap dataset for rhythm modeling. Specifically, we first crawl many rap songs, each song with both rap lyrics and audios, from the Web. For each crawled rap song, we perform a series of data preprocessing steps to extract rhythmic beats as well as beat-lyric alignment. To better model rhyme, we generate the words in a rap sentence from right to left in an autoregressive manner. Doing so we can easily identify the last few words of a sentence (now become the first words of the reverse sentence) to rhyme with. Additionally, we incorporate several rhyme-related representations into our language model to further improve the rhyming quality, and encourage the $N$-gram rhyme in generated rap lyrics through rhyme constraint during inference. We use a special token \textsc{[beat]} to represent the rhythmic beat and insert it into lyrics right before the corresponding word. In this way, we can model the beat in the lyric sequence both in training and generation. 

Inspired by the success of pre-trained language models~\cite{devlin2019bert,Radford2018GPT,yang2019xlnet,song2019mass,Yinhan2019Roberta}, we incorporate pre-training into our system. To obtain large-scale data for pre-training, we also use our data mining pipeline to collect another two datasets: 1) non-rap songs with aligned beats, which can be larger than rap dataset since non-rap songs are more general than rap songs; 2) pure lyrics, which can be even larger than non-rap songs. In the pre-training stage, we pre-train our DeepRapper model based on the above two datasets.  Then we fine-tune our pre-trained model on the rap songs with aligned beats. The fine-tuned model is used for final rap generation. Both objective and subjective evaluations verify the advantages of DeepRapper in generating rap lyrics with rhymes and rhythms.

Our main contributions can be summarized as follows:
\begin{itemize}
    \item To model rhythms in rap generation, we develop a data mining pipeline to create rap datasets with aligned rhythmic beats. 
    \item To better model rhymes, we design an autoregressive language model to generate rap lyrics from right to left with rhyme constraint. As far as we know, DeepRapper is the first to explicitly model $N$-gram rhymes.
    \item We elaborately insert the beat token inside lyrics to model the rhythmic beats. To our knowledge, DeepRapper is the first system that models rhythms for rap generation. 
\end{itemize}

\section{Background}
\begin{figure*}[ht]
	\centering
	\includegraphics[width=1.0\textwidth]{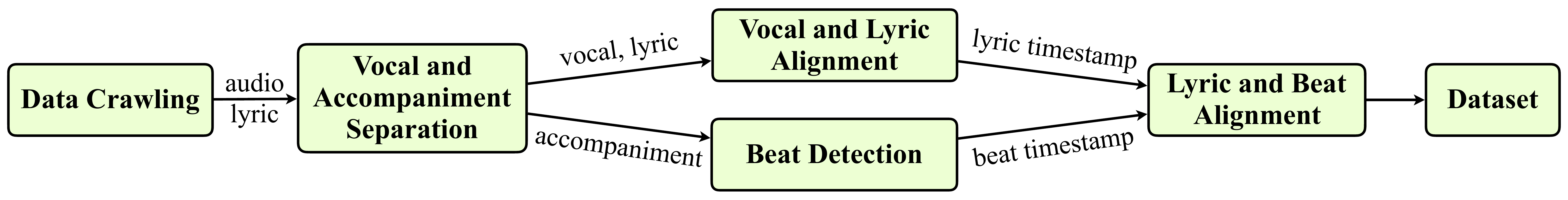}
	\caption{An overview of our data mining pipeline.}
\label{fig:data-pipeline}
\end{figure*}

Since DeepRapper generates rap lyrics with both rhyme and rhythm modeling, in this section, we briefly introduce the related background: lyric generation, rhyme modeling and rhythm modeling.

\paragraph{Lyric Generation}
Broadly speaking, lyric generation can cover rap lyric generation~\cite{potash2015ghostwriter,nikolov2020rapformer,liang2018attae}, song lyric generation~\cite{Watanabe2018melodylm,Xu2019Syllable,chen2020melodyconditioned,sheng2020songmass}, general poetry generation~\cite{zhang2014chinese,lau2018deepspeare,Piji2020SongNet,liu2020deep} and \etc. Different from previous works that leverage language model to generate lyrics similar to natural language, in this paper, we  introduce a novel language model for rap generation, with well-designed rhyme and rhythm modeling to fit the characteristics of rap lyrics. Additionally, inspired by the successes of pre-trained language models~\cite{devlin2019bert,yang2019xlnet,Yinhan2019Roberta,radford2019GPT2,song2019mass} in NLP applications, we also incorporate pre-training into our model to further improve the quality of rap generation.

\paragraph{Rhyme Modeling} 
Rhyme modeling plays an important role in rap generation, which requires the last few tokens in consecutive sentences to have the same rhyme pattern. Existing rap generation systems either directly add a special token ``$<$endLine$>$'' at the end of rap lyric to encourage the model to learn rhyme structure~\citep{potash2015ghostwriter}, or  introduce a two-step strategy for rhyme modeling that first generates rap lyrics and then adds rhyme tokens after the generated lyrics~\citep{nikolov2020rapformer}. However, these works only focused on unigram rhyme while rap appreciates more for n-gram rhyme. Although a lot of works have explored rhyme modeling in other genres, most of them cannot be directly used for rap generation. For example, poetry generation~\cite{lau2018deepspeare,zhipeng2019jiuge,liao2019gpt,Piji2020SongNet} usually used pre-defined format to control the rhyme pattern since poetry usually has fixed number of words and only cares the rhyme pattern for the last word. However, rap lyrics have diverse rhyme structures across multiple consecutive sentences and most importantly multiple consecutive words. Therefore, we introduce $N$-gram rhyme modeling in DeepRapper to handle the distinctive rhyme patterns in rap. Besides, we also train our language model in a reverse order (\ie, right to left), similar to previous works~\cite{van2020automatic}, to better model rhymes since they always occur at the end of sentence.

\paragraph{Rhythm Modeling}
Rhythm modeling is usually used in music generation~\cite{Zhu2018Xiaoice, Huang2020PopMusic,ren2020popmag} which generates the duration of notes along with the note pitch to form rhythmic beats in melody and accompaniment generation.  Different from music generation,  rap cares more about rhythmic beats instead of note pitches (i.e. melody). In this way, the generated rap lyrics need to align with the corresponding rhythmic beats in order to be rapped, otherwise it cannot be regarded as a complete rap. However, to the best of our knowledge, none of previous works have studied the rhythm modeling in rap generation. In this paper, we introduce a novel beat modeling strategy in DeepRapper for rhythm generation.


\section{Rap Dataset Mining}
Previous works~\cite{potash2015ghostwriter,liang2018attae,nikolov2020rapformer} for rap generation usually used rap datasets with only lyrics, without considering the rhythmic  beat information. To model rhythm in rap generation, the rap dataset should contain lyrics with aligned rhythmic beats. However, beat alignments are quite difficult to obtain, since their annotations require musicians with professional knowledge to identify stressing syllable in rap songs. To handle this problem, we design a data mining pipeline to automatically extract beat-lyric alignments. In this section, we introduce the details of the data mining pipeline and our mined dataset based on this pipeline.

\subsection{Data Mining Pipeline}
\label{subsec:data_pipeline}
Figure~\ref{fig:data-pipeline} overviews our data mining pipeline, which consists of 5 steps: data crawling, vocal and accompaniment separation, vocal and lyric alignment, beat detection, and lyric and beat alignment.

\paragraph{Data Crawling} 
To mine a large-scale rap dataset, we first crawl a large amount of rap songs with both lyrics and singing audios from the Web. To ensure the lyric and audio can be aligned in the sentence level which is beneficial for our later word-level beat alignment, we also crawl the start and end time of each lyric sentence corresponding to the audio. 

\paragraph{Vocal and Accompaniment Separation} 
For each rap song, we utilize \textit{Spleeter}~\cite{spleeter2020}~\footnote{https://github.com/deezer/spleeter}, a public music separation tool, to separate the vocal (containing rap singing) and accompaniment (containing rhythmic beats) from the crawled rap audio.

\paragraph{Vocal and Lyric Alignment} 
We split the separated vocals into the sentence level according to the crawled start and end time of each lyric sentence, and thus we can get the vocal-lyric alignments in the sentence level. We convert lyrics into phonemes via \textit{Phonemizer}~\footnote{https://github.com/bootphon/phonemizer} and utilize \textit{Montreal Forced Aligner}~\footnote{https://github.com/MontrealCorpusTools/Montreal-Forced-Aligner} to obtain vocal-lyric alignments in the phoneme level. Based on these phoneme-level vocal-lyric alignments, we obtain the corresponding timestamp of each word in the singing audio.

\paragraph{Beat Detection}  
To obtain the alignments between lyrics and beats, we need to know the timestamp of each beat. Therefore, we use a beat track detection tool,  \textit{Librosa}~\cite{brian_mcfee_2020_3955228}~\footnote{https://github.com/librosa/librosa}, to track the timestamp of each beat from the separated accompaniment that obtained from the second step.  

\paragraph{Lyric and Beat Alignment} 
After we obtain the timestamp of each word and each beat, we can align them together according to their timestamps. However, since a rapper may not sing a word exactly following the beat, directly using the timestamp to exactly match the word and beat is inappropriate. Therefore, we propose an approximate method to align them. Denote the word sequence of a lyric sentence as ${\rm W} = \{w_1, w_2, \cdots, w_{\left\vert {\rm W} \right\vert}\}$, and its beat sequence as ${\rm B} = \{b_1, b_2, \cdots, b_{\left\vert {\rm B} \right\vert}\}$, where $w_i$ and $b_j$ represent $i$-th word and $j$-th beat. We use $T_{w_i}$ and $T_{b_j}$ to represent the timestamps of $w_i$ and $b_j$ respectively.  For each beat $b_j$, we first filter out a word set $\tilde{\rm W} = \{w: \left\vert {T_{b_j} - T_{w}} \right\vert \leq r/2, w \in {\rm W}\}$, where $r$ represents the average duration of each word in the song (\ie, the total duration divides the number of words). Next, word $w_i$ is aligned with beat $b_j$ if it satisfies the following condition:
\begin{equation}
    w_i = \min_{w} |T_{b_j} - T_{w}|, w \in \tilde{\rm W}.
\end{equation}

\begin{table}[h]
    \centering
    \caption{The statistics of three mined datasets. The second and third column represent the number of songs and sentences for each dataset.}
    \begin{tabular}{l | r | r }
        \toprule
        \textbf{Dataset}   & \textbf{\#Songs}  & \textbf{\#Sentences} \\
        \midrule
        D-RAP   & 16,246   &  832,646      \\
        D-SONG  & 52,737   &  2,083,143     \\
        D-LYRIC  & 272,839  &  9,659,503 \\
        \bottomrule
    \end{tabular}
    \label{tab:dataset}
\end{table}

\subsection{Mined Datasets}
\begin{figure}[ht]
	\centering
	\includegraphics[width=0.95\columnwidth]{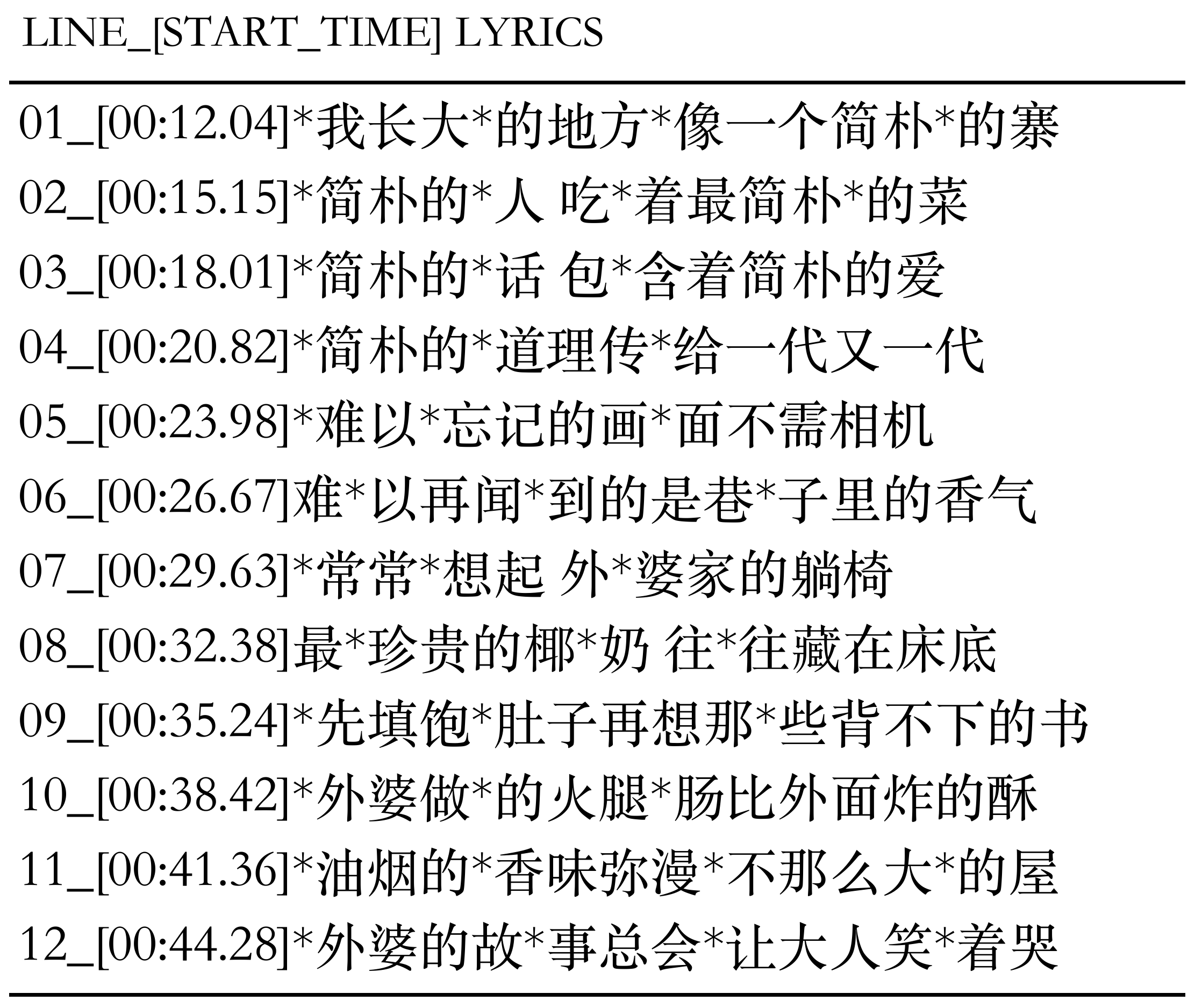}
	\caption{An example of a rap song with aligned beats in our mined ``D-RAP" dataset. `*' means a beat is aligned with the word right after `*'. Translation of the content is in supplemental materials.}
\label{fig:data-format}
\end{figure}

Using the above data mining pipeline, we obtain a rap lyric dataset with aligned beats (named as D-RAP, where D represents ``dataset''), which satisfies the requirements of building a rap generation system with both rhyme and rhythm modeling. We split the D-RAP dataset into the training and validation set with a ratio of 4:1. Since rap is only one of music genres and the number of rap songs is usually smaller compared with more general songs, we also mine another two datasets to pre-train our DeepRapper model with the same mining pipeline: 1) non-rap songs with aligned beats (named as D-SONG); 2) pure lyrics without aligned beats (named as D-LYRIC).  We summarize the statistics of the three datasets in Table~\ref{tab:dataset} and show a rap song with aligned beats from D-Rap in Figure~\ref{fig:data-format}.

\section{Rap Generation Model}
\begin{CJK*}{UTF8}{gbsn}
\begin{figure*}[ht]
	\centering
	\includegraphics[width=1.0\textwidth]{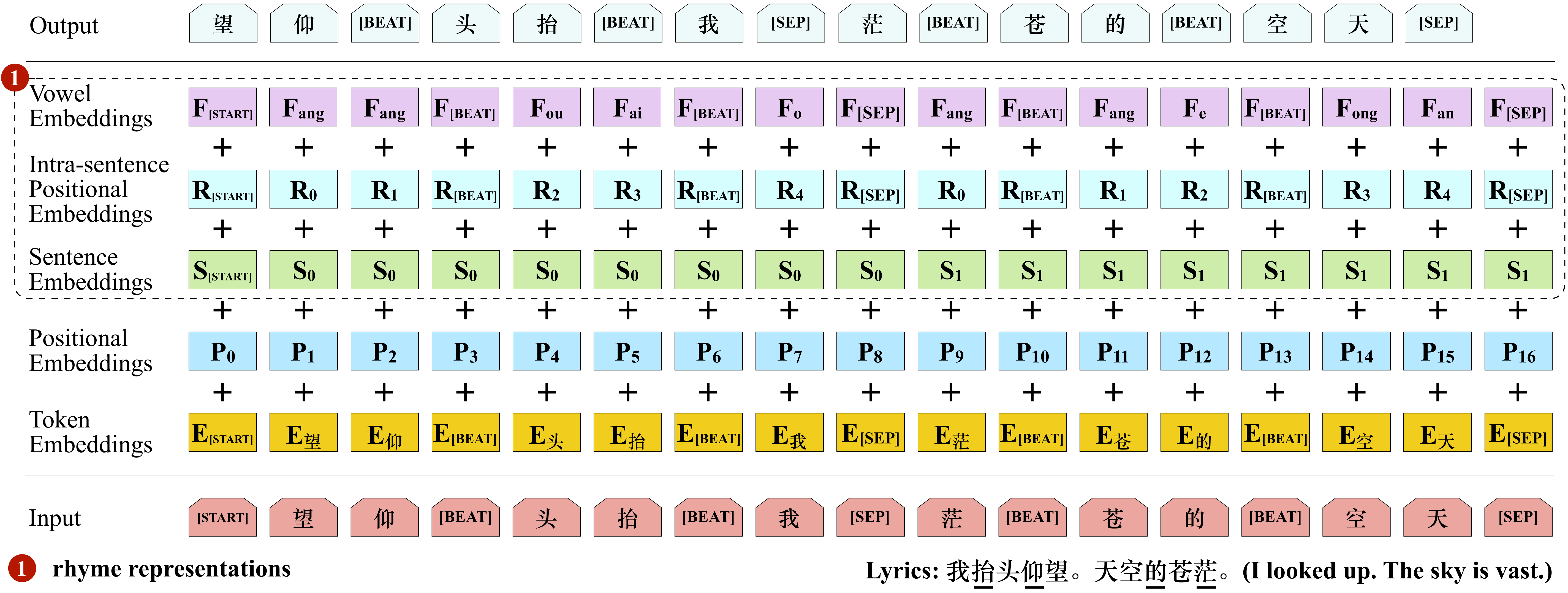}
	\caption{The architecture of the rap generation model in our \textit{DeepRapper}. The input sequence  here is a sample from a Chinese rap named《红色》(translated to English is  \emph{Red}). 
	The sample contains two lyric sentences with aligned beats. 
	Each sentence is reversed for rhyme modeling. 
	Therefore, the original form of the sample is: 
	我\underline{抬}头\underline{仰}望。天空\underline{的}苍\underline{茫}。(Translation: \emph{I looked up. The sky is vast.})
	Word with underline means that a beat is aligned with this word. Sentences are separated by special token `[SEP]'. Token `[START]' represents the start of a song. }
\label{fig:gpt-model}
\end{figure*}
\end{CJK*}

In this section, we introduce the architecture of our rap generation model, and the details of its rhyme modeling and rhythm modeling.     

\subsection{Model Overview}
Figure~\ref{fig:gpt-model} illustrates the detailed architecture of our rap generation model. We use Transformer~\cite{Vaswani2017transformer} to build an autoregressive language model~\cite{Radford2018GPT,radford2019GPT2} for rap generation, and introduce several new designs: 1) To better model rhymes, our model generates a sentence from right to left, since rhyming words are always at the end of the sentence; 2) As aforementioned, rhythms are critical for rap performance, so we insert a special token \textsc{[beat]} for explicit beat modeling; 3) Unlike original Transformer with only word embedding and positional embedding, we add multiple additional embeddings to better model rhymes and rhythms. Next, we introduce our rhyme modeling in subsection~\ref{subsubsec:rhyme} and rhythm modeling in subsection~\ref{subsubsec:beat}.

\subsection{Rhyme Modeling}
\label{subsubsec:rhyme}
Rhymes are the key to form a good rap flow. In DeepRapper, we model rhymes with three components: 1) reverse-order language model; 2) rhyme representation; and 3) rhyme constraint.

\subsubsection{Reverse-Order Language Model} 
\begin{figure}[!t]
	\centering
	\includegraphics[width=0.95\columnwidth]{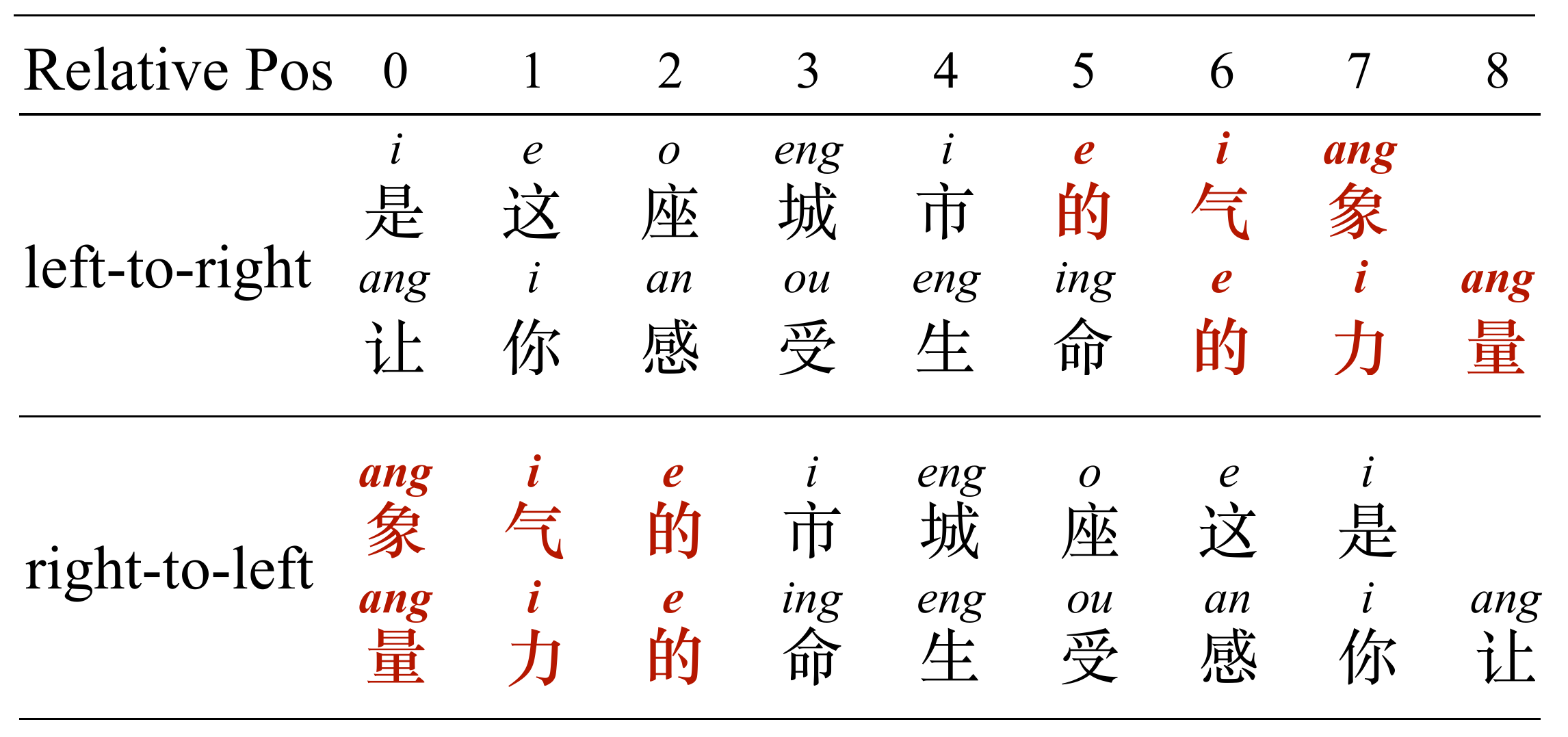}
	\caption{Comparison between sentences in left-to-right and right-to-left order. The superscript of each token represents  its rhyme. }
\label{fig:example-rhyme}
\end{figure}

Rhyming words usually occur at the end of each lyric sentence. If using a standard autoregressive language model and generating tokens from left to right, we need to identify whether the current generation step is the end of a sentence, which decides whether to generate rhyming words to be consistent with that in previous sentences. Therefore, to better model rhymes, we use a reverse-order language model to generate sentences from right to left, as shown in Figure~\ref{fig:gpt-model}. Doing so we can easily identify the last few words of a sentence (now become the first few words of the reverse sentence) to control their rhymes. Note that we only reverse words inside a sentence, and still generate different sentences in the original order. Figure~\ref{fig:example-rhyme} compares the sentences in left-to-right order and right-to-left order, from which we can see that rhyming words of each sentence share the same relative positions (offset to the first token) in the reverse order, and are easy to model and control. 

\subsubsection{Rhyme Representation}
Rhyming words have two important features: 1) its vowel that used for rhyming and 2) its relative position in a sentence to decide the correspondence between the rhyming words in consecutive sentences (e.g., in the reverse order setting, the first/second word of the current sentence should be rhymed with the first/second word in the previous sentence). 

We use the vowel in the Pinyin~\footnote{Pinyin is the standard phoneme for Chinese.} of Chinese characters to represent their rhymes. To this end, we build a vowel dictionary $\mathcal{F}(\cdot)$ to identify the vowel of each word. As shown in Figure~\ref{fig:gpt-model}, we add an additional vowel embedding $\mathbf{F}$ and an intra-sentence relative positional embedding $\mathbf{R}$ to enhance rhyme representation for each token. Besides, to better identify different sentences, we introduce a sentence embedding $\mathbf{S}$ to differentiate different sentences. 

\subsubsection{Rhyme Constraint}
In addition to reverse-order language model and rhyme representation, we also introduce rhyme constraint to improve the quality of rhyme generation in inference. As shown in Figure~\ref{fig:example-rhyme}, sentences in rap lyrics not only rhyme with the last token, but also with multiple consecutive tokens at the end. We call this phenomenon as $N$-gram rhymes, which mean the current sentence and the previous sentence keep the same rhyme for the last $N$ consecutive tokens. To our knowledge, no previous work has investigated $N$-gram rhymes ($N > 1$), although it is important to improve rap quality. Our proposed rhyme constraint enables our model to adjust the probability of next predicted token to further encourage $N$-gram rhyme generation. The constraint is introduced as follows. 

To generate the $i$-th word $w_i$ in the standard inference procedure, we usually choose the predicted token with the maximum probability, \ie, $w_i = \arg\max p(w|w_{<i};\theta)$, where $w_{<i}$ denotes the words before position $i$ in the reverse sentence and $\theta$ is the model. When the words before position $i$ of the current and previous sentence have the same rhyme pattern, we will use an adjusted probability distribution $\tilde{p}(w|w_{<i};\theta)$ to encourage the $i$-th generated word to be rhymed according to the $i$-th word in the previous sentence, so as to form $N$-gram rhymes. The adjusted probability distribution  $\tilde{p}(w|w_{<i};\theta)$ is:
\begin{equation} \label{eq2}
    \tilde{p}(w|w_{<i};\theta)
    = \alpha \cdot p(w|w_{<i};\theta) + (1-\alpha) \cdot \pi(w)
\end{equation}
where $\pi(w)$ is a vowel check function and $\alpha$ is a hyper-parameter to balance the two terms. Here, $\pi(w)$ is 1 if the predicted $w$ has the same vowel with the $i$-th token in the previous sentence, otherwise 0. In other words, when predicting $i$-th token ($i \leq N$), we encourage our model to pay more attention for these words with same vowel with the $i$-th token in the previous sentence. In this way, the model tends to generate $N$-gram rhymes with large $N$.

\subsection{Rhythm Modeling}
\label{subsubsec:beat}
Generating lyrics with aligned beats is necessary since rap lyrics need to be rapped with rhythmic beats. Therefore, we model and generate rhythmic beats along with the lyrics with a specific symbol: we regard beat as a special token \textsc{[beat]} and insert it into lyric sequences for model training. As shown in Figure~\ref{fig:gpt-model}, we insert \textsc{[beat]} before its aligned words like the following examples:
\begin{CJK*}{UTF8}{gbsn}
``我\textsc{[beat]}抬头\textsc{[beat]}仰望。天空\textsc{[beat]}的苍\textsc{[beat]}茫。".
\end{CJK*}

\begin{figure}[!t]
	\centering
    \includegraphics[width=0.95\columnwidth]{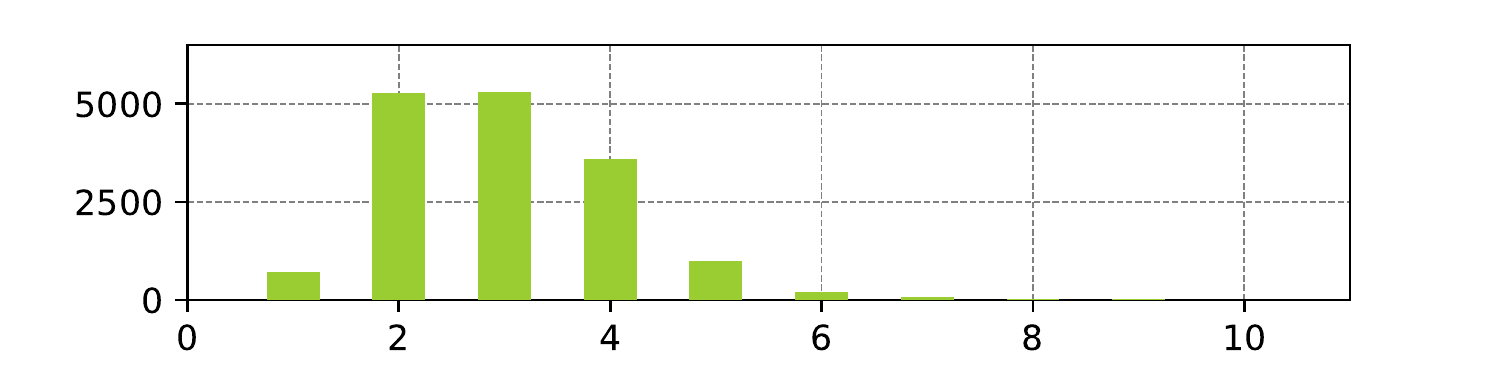}
	\caption{The distribution of beat frequencies in D-RAP. X-axis: beat frequency (the ratio between the total number of words and the total number of beats in a rap song). Y-axis: the number of songs.}
\label{fig:beat-frequency}
\end{figure}

Rap usually contains different beat frequencies, i.e., the ratios between the total number of words and the total number of beats in a rap song. To explicitly model and generate rap with different beat frequencies, we use three tokens \textsc{[s]}, \textsc{[m]}, and \textsc{[f]} to represent the slow, medium and fast beat frequencies and add the corresponding tokens at the start of a rap song for training and inference. In our D-RAP dataset, the distribution of beat frequency is displayed in Figure~\ref{fig:beat-frequency}. According to the distribution, we assign \textsc{[s]}, \textsc{[m]}, and \textsc{[f]} to songs with beat frequency less than 3, equal to 3, and greater than 3 respectively.

\section{Experimental Setup}
\subsection{Model, Data, and Training Configuration}
Our DeepRapper model is built on the autoregressive Transformer decoder~\cite{Vaswani2017transformer,Radford2018GPT,radford2019GPT2}, where the hidden size, the number of attention heads and the number of Transformer layers are set as 768, 12, 12. The dimension of all different kinds of embedding in DeepRapper is set as 768.
Considering there is no existing pre-trained language model in reverse order, we do not utilize any pre-trained language models for initialization. Instead, we  first pre-train our model on D-LYRIC and D-SONG for 2 millions steps, and then fine-tune our model on D-RAP with 3K steps as the size of D-RAP is smaller than our pre-training corpus.
We convert each song to a sequence with a length of $1024$ tokens by cutting longer sequence or padding shorter sequence. Our model is trained with a batch size of 8 songs on 4 NVIDIA TITAN V GPUs. We use Adam optimizer with a learning rate of $0.00015$, $\beta_1 = 0.9$, $\beta_2=0.999$, and $\epsilon=10^{-6}$. We set the maximum value of $N$-gram rhyme as 3 and the hyper-parameter $\alpha$ in Equation~\ref{eq2} as 0.95. Samples are generated conditioned on a given sentence in reference.


\begin{table*}[htb]
    \centering
    \caption{Results of objective and subjective evaluations. ``+PT" means using pre-training. Since the two baselines do not include beat information, we only compare in perplexity (PPL), rhyme accuracy (RA) and rhyme density (RD) for objective evaluation. For subjective evaluation, we report the average annotation score of theme, fluency, rhyme quality, and rhyme diversity.}
    \begin{tabular}{l| r | r | r | r | r | r | r  }
    \toprule
    \multirow{2}{*}{\textbf{Model}} & \multicolumn{3}{c|}{\textbf{Objective Evaluation}} & \multicolumn{4}{c}{\textbf{Subjective Evaluation}} \\
\cline{2-4} \cline{5-8}
    & \textbf{PPL} $\downarrow$ & \textbf{RA} $\uparrow$ & \textbf{RD} $\uparrow$ & \textbf{Theme} $\uparrow$ & \textbf{Fluency} $\uparrow$ & \textbf{Quality} $\uparrow$ & \textbf{Diversity} $\uparrow$ \\
    \midrule
    Baseline & 24.65 & 32.29 & 0.23 & 3.13 & 2.55    & 3.10  & 3.46 \\
    Baseline + PT & 13.47 & 39.59 & 0.35 &  3.41 & 2.69 & 3.25 & 3.65 \\
    DeepRapper & \textbf{5.65} & \textbf{66.95} & \textbf{1.65} & \textbf{3.67}	& \textbf{3.57}	   & \textbf{4.16}	 & \textbf{4.14} \\
    \bottomrule
    \end{tabular}
    \label{tab:objective_evaluation}
\end{table*}

\subsection{Evaluation Metrics}
\label{subsec:evaluation_metrics}
In this subsection, we introduce the objective and subjective metrics to evaluate the quality of the generated raps.

\paragraph{Objective Evaluation} We evaluate the generated raps in terms of the quality of language, rhyme and rhythm. We choose five metrics to evaluate our model: 1) \textbf{Perplexity} (PPL), a standard metric to evaluate the quality of a language model; 2) \textbf{Rhyme Accuracy} (RA), the ratio of sentences that have correctly predicted rhymes; 3) \textbf{Rhyme Density} (RD),  the longest rhyme of a song, averaged over all songs, which is introduced by \citet{malmi2016dopelearning} to measure the quality of rhyming fluency; 4) \textbf{Combo-$\mathbf{N}$}, the maximum number of consecutive sentences with the same $N$-gram rhyme in a rap song, averaged over all songs, where we study $N=1,2,3$; 5) \textbf{Beat Accuracy} (BA), the accuracy of our model in beat prediction, under the teacher-forcing mode.

\paragraph{Subjective Evaluation} Similar to previous works~\cite{zhang2014chinese,nikolov2020rapformer} in artistic creation, we also use human evaluation to accurately evaluate the quality of the generated raps. We invite 10 participants with professional knowledge in music as human annotators to evaluate 100 sampled raps. Each annotator is required to score from 1 (Poor) to 5 (Perfect) on the following perspectives: 1) the clearness of the theme of the rap lyrics; 2) the fluency of the rap lyrics; 3) the quality of the rhyme; 4) the diversity of the rhyme. The averaged score of all annotators on all sampled raps is used as the evaluation score for each perspective. 

\section{Experimental Results}
\paragraph{Results}
Table~\ref{tab:objective_evaluation} shows the objective and subjective results of DeepRapper compared with two baselines: 1) Baseline: a standard autoregressive language model with the same model configuration with DeepRapper but without our proposed rhyme and rhythm modeling; 2) Baseline + PT, using pre-training on Baseline. We have several observations from Table~\ref{tab:objective_evaluation}: 1) DeepRapper achieves better perplexity, rhyme accuracy and rhyme density than the two baselines, which demonstrates the advantages of our method in generating high-quality rap lyrics with accurate and diverse rhymes. 2) DeepRapper achieves better scores in all subjective metrics, demonstrating that DeepRapper can generate high-quality and rhyming raps that accord with human taste. 3) Pre-training improves the performance of baseline in both objective and subjective metrics, which indicates the importance of pre-training. However, its performance is still worse than DeepRapper. 

\begin{table}[h]
\centering
\caption{The ablation studies on each component in DeepRapper. ``-" means removing the corresponding component. ``Rhyme", ``Rhythm" and ``PT" represent rhyme modeling, rhythm modeling and pre-training. ``RO", ``VE", ``IPE" and ``SE" mean reverse-order, vowel embedding, intra-sentence position embedding and sentence embedding. }
\begin{tabular}{l | c | c | c | c  }
\toprule
\textbf{Model}  & \textbf{PPL} $\downarrow$  & \textbf{RA} $\uparrow$ & \textbf{BA} $\uparrow$  & \textbf{{RD}} $\uparrow$ \\
\midrule
{DeepRapper}                       & 5.65  & 66.95  & 79.00  & 1.65  \\
\midrule

{- Rhyme}    & 7.13  & 41.25  & 79.42  & 0.35   \\
{~~~~~- RO}  & 3.70  & 44.58  & 82.83  & 0.41   \\   
{~~~~~- VE}  & 6.06  & 66.11  & 78.19  & 1.46   \\
{~~~~~- IPE} & 5.81  & 63.86  & 77.29  & 1.50    \\
{~~~~~- SE}  & 6.69  & 66.52  & 80.37  & 1.63   \\

{- Rhythm}  & 4.66  & 66.08  & --     & 1.56  \\
{- PT}      & 28.69 & 40.77  & 76.40  & 1.98 \\

\bottomrule
\end{tabular}
\label{tab:ppl}
\end{table}

\paragraph{Ablation Studies}
To further validate the necessity of each component in DeepRapper, we conduct a series of ablation studies, including removing rhyme modeling, rhythm modeling and pre-training, respectively. The results are reported in Table~\ref{tab:ppl}. We have several observations: 1) Removing rhyme modeling affects rhyme quality a lot as it results in a dramatic drop in rhyme accuracy and rhyme density; 2) Removing each specific design in rhyme modeling (i.e., RO: reverse order language model, VE: vowel embedding, IPE: intra-sentence position embedding, SE: sentence embedding) causes worse rhyme accuracy and rhyme density. Specifically, while removing RO leads to a better PPL since left-to-right order can be more easily modeled than right-to-left order according to the analysis in~\citet{wu2018beyond}, it causes large accuracy drop in rhyme quality. 3) Apparently, DeepRapper without rhythm modeling cannot produce any beat information; 4) DeepRapper without pre-training affects the perplexity and rhyme accuracy a lot, however, obtains a higher rhyme density. 
The reason is that without pre-training, DeepRapper tends to copy previous rhyme tokens due to the lack of generalization (larger PPL). To verify this, we count the repetitive rate of rhyming words and found that the rate of DeepRapper is $23.8\%$ while without pre-training is $42.5\%$, which is higher than using pre-training. The above results verify the effectiveness of each component in DeepRapper.

\paragraph{$N$-gram Rhyme}
To highlight the advantage of DeepRapper in modeling ${\rm N}$-gram rhyme, we use Combo-${\rm N}$ to measure the ability of each design in DeepRapper to model ${\rm N}$-gram rhyme. The results are reported in Table~\ref{tab:n_gram}. We can find that 1) The model without rhyme modeling can hardly generate good rhyme, regardless of the value of $\rm N$ in ${\rm N}$-gram; 2) Removing rhyme constraint also weakens the capacity of generating ${\rm N}$-gram rhyme. These results further demonstrate the importance of our rhyme modeling and rhyme constraint in generating multiple consecutive rhymes. 

\begin{table}[ht]
\centering
\caption{Quality of $N$-gram rhyme in terms of Combo-${\rm N}$ . ``- Rhyme" means removing rhyme modeling and ``- RC" means removing rhyme constraint during inference.}
    \small
    \begin{tabular}{l|r | r | r}
    \toprule
    Model & \textbf{{Combo-1}} & \textbf{{Combo-2}} & \textbf{{Combo-3}} \\
    \midrule
    DeepRapper & \textbf{87.10} & \textbf{18.16} & \textbf{9.10} \\
    \midrule
    {- Rhyme}  & 7.37 & 2.85 & 1.71 \\
    {- RC}     & 70.08 & 13.73 & 3.77 \\
    \bottomrule
    \end{tabular}
    \label{tab:n_gram}
\end{table}

\paragraph{Beat Frequency}
To better measure the beat quality, we randomly generate about 5,000 samples by DeepRapper and DeepRapper with beat frequency control. We propose the First Order Distribution (FOD) and the Second Order Distribution (SOD) and measure the distance (via Wasserstein Distance~\cite{vallender1974calculation}) of these distributions between the generated samples and our DRAP dataset. We define the interval of the current \textsc{[BEAT]} as the number of words between the current \textsc{[BEAT]} and the next \textsc{[BEAT]}. Therefore, the FOD is defined as the distribution of the interval of the current \textsc{[BEAT]}. Similarly, the SOD is defined the distribution of the difference between the interval of the current \textsc{[BEAT]} and the next \textsc{[BEAT]}. The results of the distance are normalized into $[0, 1]$ and are reported in Table~\ref{tab:beat}. It can be seen that DeepRapper with beat frequency control achieves better performance in beat modeling, which indicates the importance of beat frequency control in beat modeling.


\begin{table}[ht]
\centering
\caption{Measurement of beat generation. ``+ Beat Frequency'' represents DeepRapper with beat frequency control.}
    \begin{tabular}{l| c | c}
    \toprule
    Model & FOD & SOD \\
    \midrule
    DeepRapper & 0.1107 & 0.0514 \\
    \midrule
    {+ Beat Frequency}  & \textbf{0.1091} & \textbf{0.0502} \\
    \bottomrule
    \end{tabular}
    \label{tab:beat}
\end{table}

\paragraph{Case Analyses on Generated Raps}
\begin{figure}[!b]
	\centering
	\includegraphics[width=0.95\columnwidth]{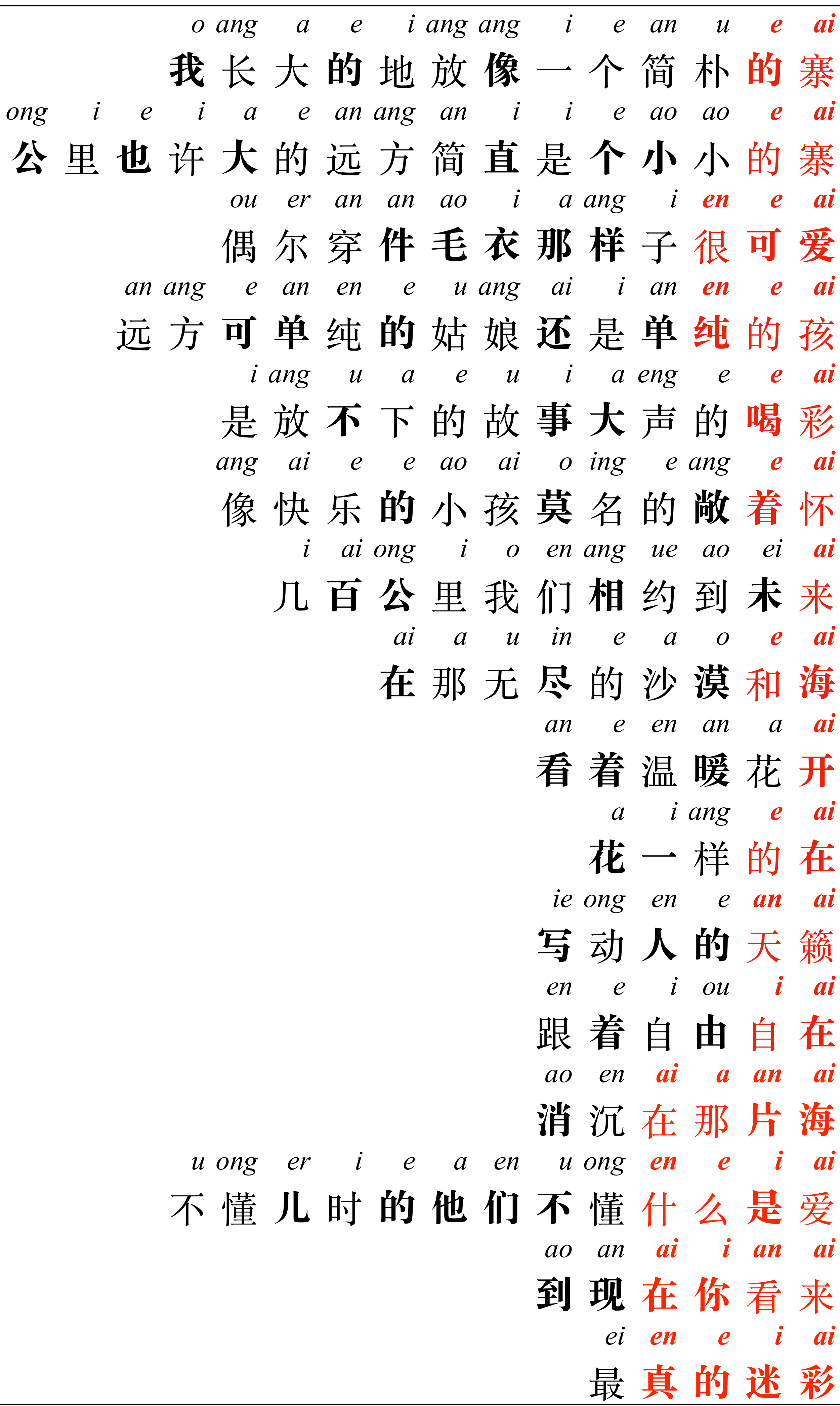}
	\caption{A rap generated by \textit{DeepRapper}. For each example, we provide the corresponding vowel of each word. Vowels in red color represents that the word rhymes with previous sentence. Bold word means a beat is aligned with the word. The translation of the example is attached in supplemental materials.}
	\vspace{-15pt}
\label{fig:case-study}
\end{figure}
We list a sample case from our generated raps in Figure~~\ref{fig:case-study} to demonstrate the good quality of the raps generated by DeepRapper. The sample is generated by feeding the first sentence of the example in Figure~\ref{fig:data-format} to DeepRapper. As we can see, the generated sample exhibits good theme, fluency and rhyme. The sample is a rap with a number of 1-gram, 2-gram, 3-gram, and even 4-gram rhyme. The generated lyrics depicts the fond memories of childhood and the beautiful visions for the futures. 
We also provide a group of samples generated with beat frequency control. To save space, we put them and the translation of all the samples to Appendix. More samples are provided in \href{https://deeprapper.github.io}{https://deeprapper.github.io}.

\section{Conclusions}
In this paper, we develop \textit{DeepRapper}, a novel Transformer-based rap generation system, which leverages rhyme modeling, rhythm modeling and pre-training for rap generation. Considering there is no available rap dataset with aligned rhythmic beats for rhythm modeling, we propose a data mining pipeline to mine a rap dataset with beat-lyric alignments. We leverage right-to-left generation, rhyme representation and rhyme constraint to better model rhyme and encourage ${\rm N}$-gram rhyme, and explicitly model beat information by insert beat token beside the corresponding word in the lyric sequence. To our knowledge, DeepRapper is the first system to generate rap with both rhymes and rhythms. Both objective and subjective evaluations demonstrate that DeepRapper generates  high-quality raps with good rhymes and rhythms. Thanks to the design of DeepRapper, we can further build another rap singing system to sing out the raps according to the rhymes and rhythms, which we leave as future work. We also leave Multilingual \textit{DeepRapper} as future work.

\section*{Acknowledgements}
We would like to acknowledge the anonymous reviewers for their insightful comments. 
Research on this paper was supported by Hong Kong Research Grants Council under grant 16204920.

\section*{Ethical Considerations}
The proposed framework can be considered a novel language model for rap generation in automatic artistic creation. Specifically, the proposed framework has been configured with novel rhyme modeling as rhyme is quite important in music genres. Therefore, our proposed framework is also beneficial for generating other music genres. On the other hand, although we collect large-scale lyric data for pre-training, it still cannot fully utilize the potential of pre-training. In the future, we expect to employ more large-scale data in the open domain plus the music domain for pre-training to improve the capacity of the language model. In addition, our training datasets may have biases, which may bring some potential risks of model bias. Hence, we encourage future works to study how to apply other techniques in mitigating similar problems in our framework.

\bibliographystyle{acl_natbib}
\bibliography{acl2021}

\appendix

\section{Comparison with GhostWriter}
We provide a comparison between DeepRapper and GhosterWriter~\cite{potash2015ghostwriter} in Table~\ref{tab:ghostwriter}. The results show that both DeepRapper and baselines outperform GhosterWriter in terms of PPL, rhyme accuracy, and rhyme density on rap generation tasks.
\begin{table}[htb]
    \centering
    \caption{The comparison of DeepRapper with GhosterWriter in perplexity (PPL), rhyme accuracy (RA) and rhyme density (RD). ``+PT" means using pre-training.}
    \begin{tabular}{l| r | r | r }
    \toprule
    \multirow{2}{*}{\textbf{Model}} & \multicolumn{3}{c}{\textbf{Objective Evaluation}} \\
\cline{2-4}
    & \textbf{PPL} $\downarrow$ & \textbf{RA} $\uparrow$ & \textbf{RD} $\uparrow$  \\
    \midrule
    GhostWriter & 58.95 & 7.28 & 0.17 \\
    Baseline & 24.65 & 32.29 & 0.23 \\
    Baseline + PT & 13.47 & 39.59 & 0.35 \\
    DeepRapper & \textbf{5.65} & \textbf{66.95} & \textbf{1.65}  \\
    \bottomrule
    \end{tabular}
    \label{tab:ghostwriter}
\end{table}

\begin{figure}[htb]
	\centering
	\includegraphics[width=\columnwidth]{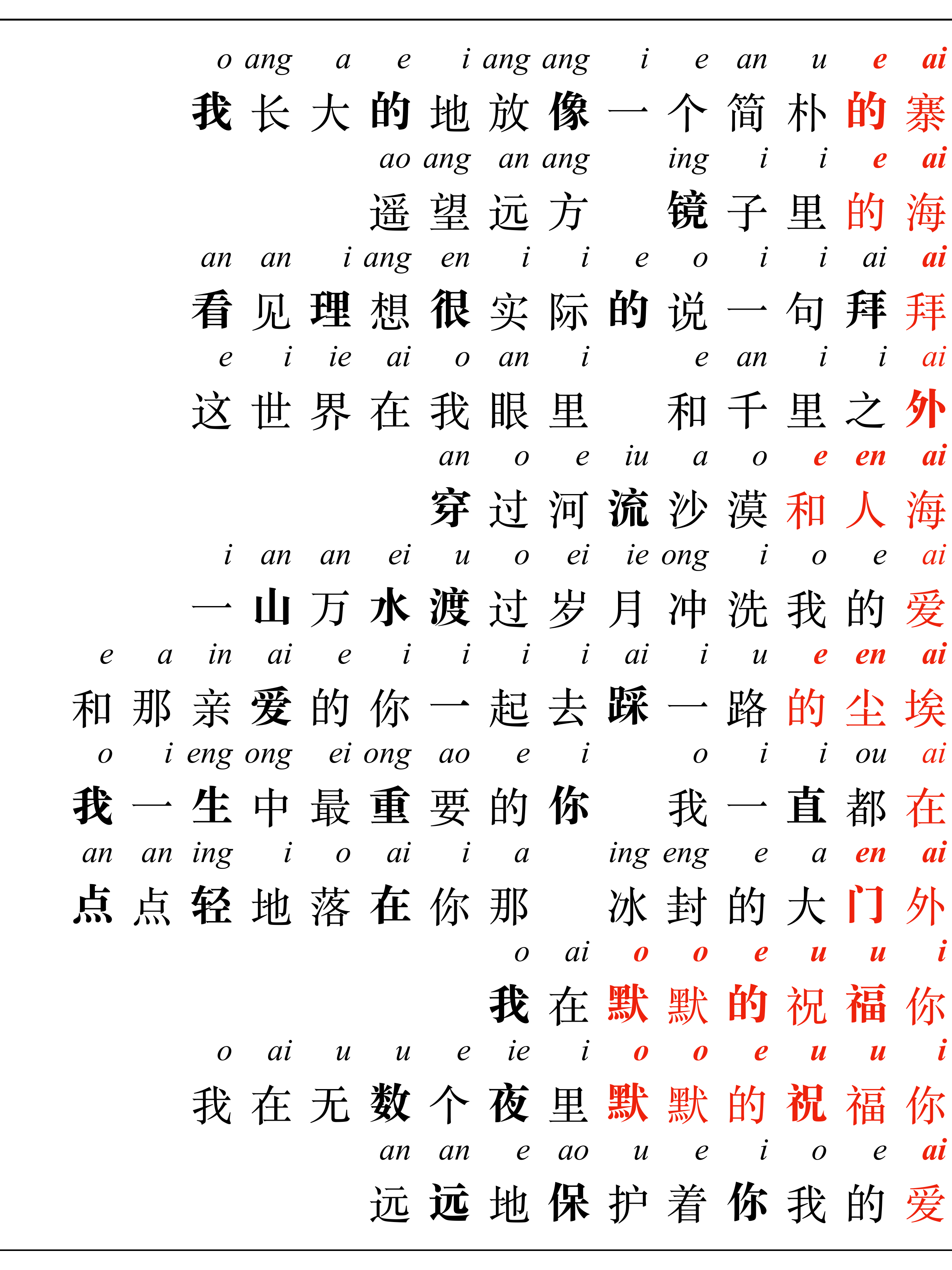}
	\caption{Rap generated of fast beat frequency. Vowels in red color represents that the word rhymes with previous sentence. Bold word means a beat is aligned with the word. }
\label{fig:bfc-fast}
\end{figure}

\section{Samples with Beat Frequency Control}\label{app:bfc}
\paragraph{Fast} Figure~\ref{fig:bfc-fast} provides a rap generated by DeepRapper with fast beat frequency, which the frequency is 4.3. The rap express ones beat wished to his/her lover. The following is the translation of texts in Figure~\ref{fig:bfc-fast}.

\begin{CJK*}{UTF8}{gbsn}
\begin{table}[H]
\begin{tabularx}{\columnwidth}{l}
我长大的地方像一个简朴的寨 \\
The place where I grew up is like a simple\\ village \\
遥望远方 镜子里的海 \\
Looking into the distance, the sea is in the\\ mirror \\
看见理想很实际的说一句拜拜 \\
See my dream and say goodbye \\
这世界在我眼里 和千里之外 \\
The world is sometimes in my eyes and \\
sometimes thousands of miles away \\
穿过河流沙漠和人海 \\
Across rivers, deserts and crowds \\
一山万水渡过岁月冲洗我的爱 \\
A mountain and a million rivers wash my love \\
through the years \\
和那亲爱的你一起去踩一路的尘埃  \\
Step on the dust all the way with dear you \\
我一生中最重要的你 我壹一都在  \\
The most important you in my life, I'll always \\
by your side \\
点点轻地落在你那 冰封的大门外 \\
Little by little, it falls outside your frozen gate \\
我在默默的祝福你 \\
I am blessing you silently \\
我在无数个夜里默默地祝福你 \\
I have secretly blessed you for countless nights \\
远远地保护着你我的爱 \\
Protecting you and my love from a distance \\
\end{tabularx}
\end{table}    

\begin{figure}[htb]
	\centering
	\includegraphics[width=\columnwidth]{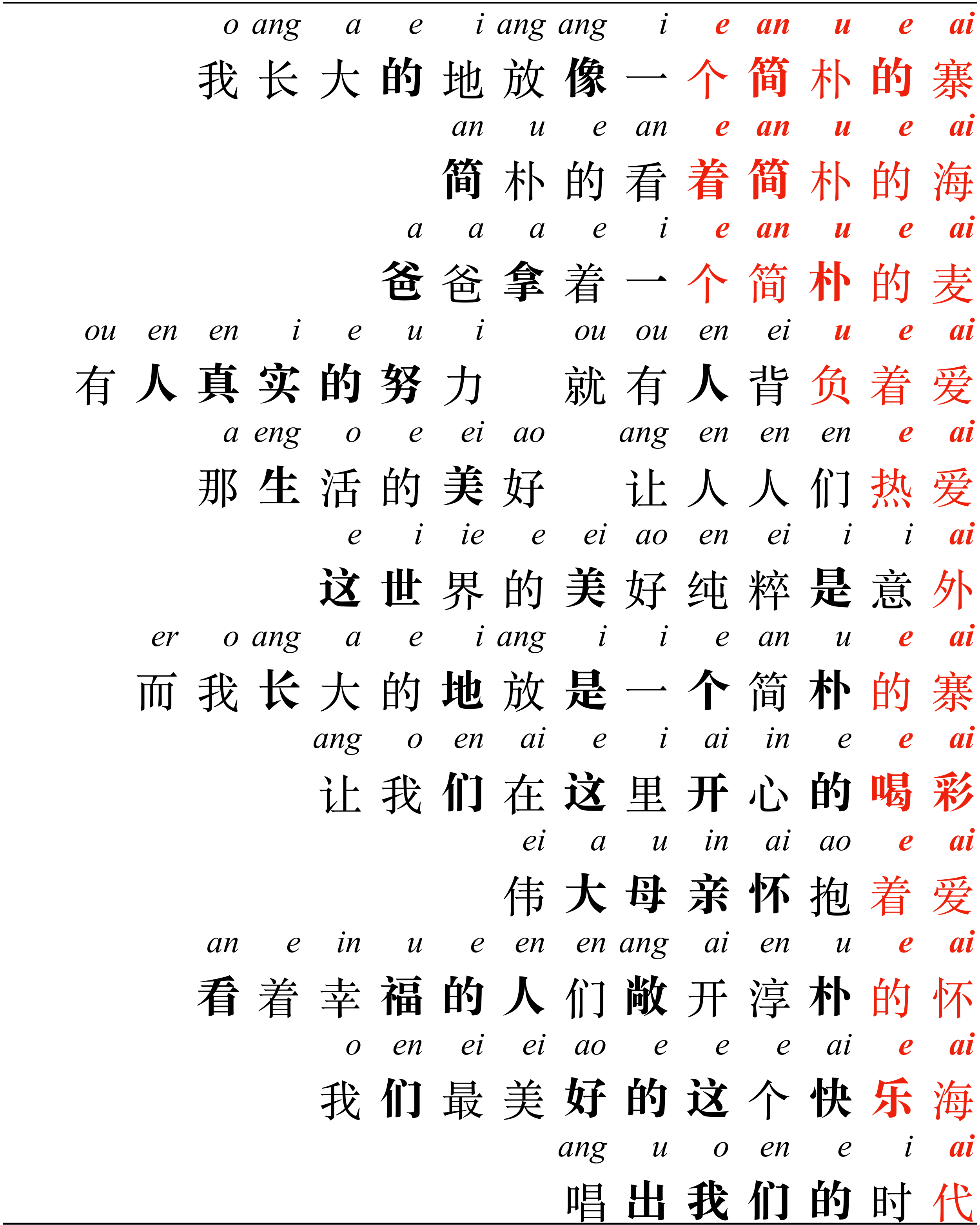}
	\vspace{-5pt}
	\caption{Rap generated of medium beat frequency. Vowels in red color represents that the word rhymes with previous sentence. Bold word means a beat is aligned with the word. }
\label{fig:bfc-medium}
\end{figure}

\paragraph{Medium} Figure~\ref{fig:bfc-medium} provides a rap generated by DeepRapper with medium beat frequency, which the frequency is 2.6. The rap  praises the times we live in. The following is the translation of texts in Figure ~\ref{fig:bfc-medium}.

\begin{table}[H]
\begin{tabularx}{\columnwidth}{l}
我长大的地方像一个简朴的寨 \\
The place where I grew up is like a simple\\ 
village \\
简朴的看着简朴的海 \\
Simply looking at the simple sea \\
爸爸拿着一个简朴的麦 \\
Dad holding a simple wheat \\
有人真实的努力 就有人背负着爱 \\
Someone takes effort, somebody is carrying love \\
那生活的美好 让人人们热爱 \\
The beauty of life makes people love \\
这世界的美好纯粹是意外 \\
The beauty of this world is pure accident \\
而我长大的地方是个简朴的寨 \\
And the place where I grew up is a simple\\
village \\
让我们在这里开心的喝彩 \\
Let's cheer happily here \\
伟大母亲 怀抱着爱 \\
Great mother embrace love \\
看着幸福的人们敞开淳朴的怀 \\
Watching happy people open their simple arms \\
我们最美好的这个快乐海 \\
We are in the most beautiful happy sea \\
唱出我们的时代 \\
Sing our time \\
\end{tabularx}
\end{table} 

\begin{figure}[htb]
	\centering
	\includegraphics[width=\columnwidth]{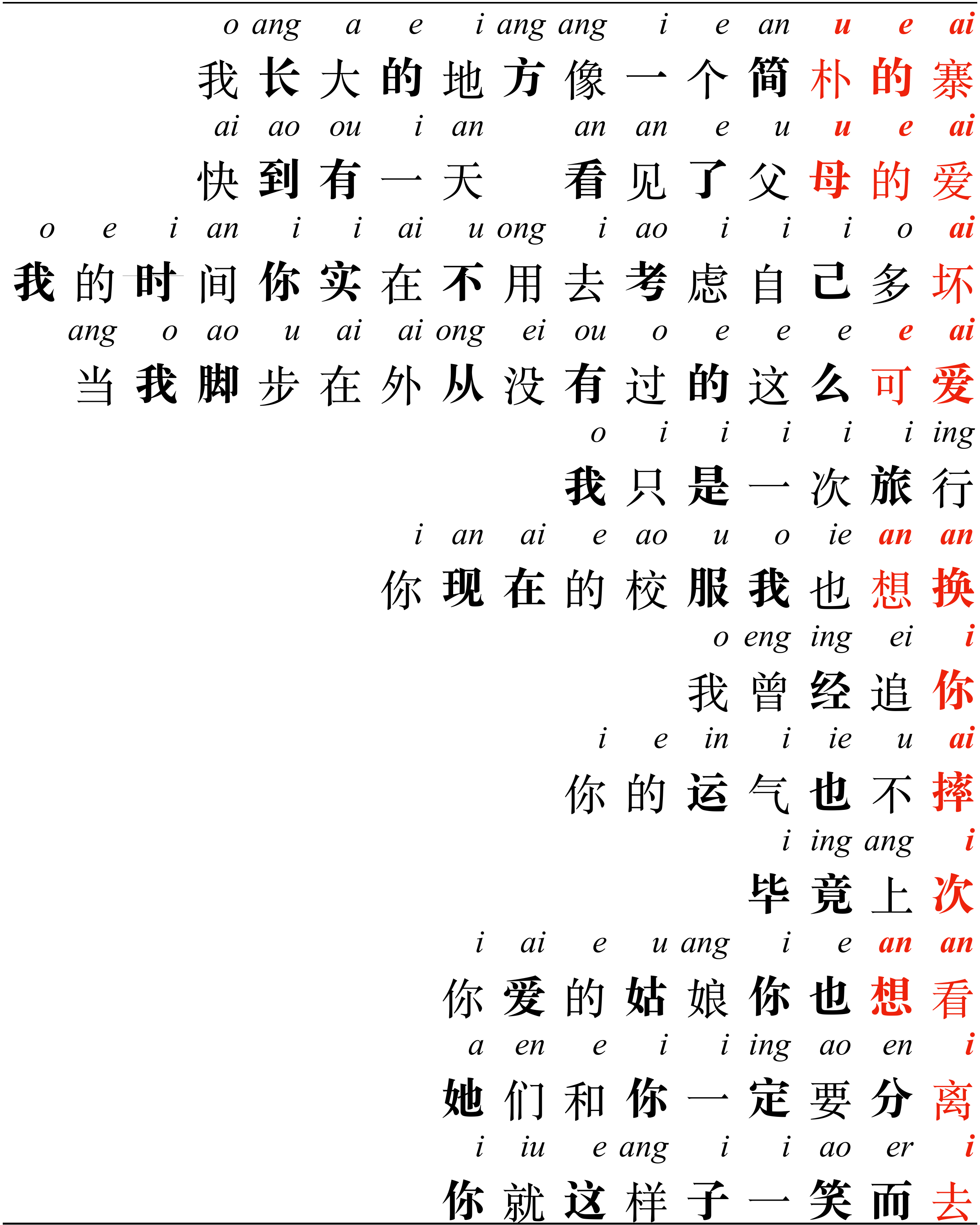}
	\caption{Rap generated of slow beat frequency. Vowels in red color represents that the word rhymes with previous sentence. Bold word means a beat is aligned with the word. }
\label{fig:bfc-slow}
\end{figure}

\paragraph{Slow} Figure~\ref{fig:bfc-slow} provides a rap generated by DeepRapper with slow beat frequency, where the frequency is 2.1. The rap express ones relief from life. The following is the translation of texts in Figure~\ref{fig:bfc-slow}.

\begin{table}[H]
\begin{tabularx}{\columnwidth}{l}
我长大的地方像一个简朴的寨 \\
The place where I grew up is like a simple\\ 
village \\
快到有一天 看见了父母的爱 \\
Almost one day I saw the love of my parents \\
我的时间你实在不用去考虑自己多坏 \\
You don't have to think about how bad \\
you are in my time \\
当我脚步在外从没过的这么可爱 \\
I've never been so cute when I'm out \\
我只是一次旅行 \\
I'm just a trip to your life \\
你现在的校服我也想换 \\
I want to change your current school uniform \\
我曾经追你 \\
I used to chase you \\
你的运气也不摔 \\
Your luck won't fall \\
毕竟上次 \\
After all last time \\
\end{tabularx}
\end{table}

\begin{table}[H]
\begin{tabularx}{\columnwidth}{l}
你爱的姑娘你也想看 \\
You want to see the girl you love \\
她们和你一定要分离 \\
They must be separated from you \\
你就这样子一笑而去 \\
You just leave with a smile \\
\end{tabularx}
\end{table} 
\end{CJK*}

\section{Translation of Chinese Examples in the Paper}\label{app:translation}
Words in red are rhymes.
\begin{CJK*}{UTF8}{gbsn}
\paragraph{Translation of Chinese in Figure~2}~
\begin{table}[H]
\begin{tabularx}{\columnwidth}{l}
    我长大的地方像一\color{red}{个简朴的寨} \\
    The place where I grew up is like a \\ simple village \\
    简朴的人 吃着最\color{red}{简朴的菜} \\
    Simple people eat the simplest dishes \\
    简朴的话 包含\color{red}{着简朴的爱} \\
    Simple words contain simple love \\
    简朴的道理传给一代又一\color{red}{代} \\
    Simple principles are passed on from generation \\to generation \\
\end{tabularx}
\end{table}

\begin{table}[H]
\begin{tabularx}{\columnwidth}{l}
    难以忘记的画面不需\color{red}相机 \\
    Unforgettable picture do not need camera \\ to capture\\
    难以再闻到的是巷子里\color{red}的香气 \\
    What is hard to smell is the aroma in the alley \\
    常常想起 外婆家\color{red}的躺椅 \\
    I often think of grandma’s recliner \\
    最珍贵的椰奶 往往藏在\color{red}床底 \\
    The most precious coconut milk is often hidden \\ under the bed \\
    先填饱肚子再想那些背不\color{red}下的书 \\
    Fill your stomach first, then think about the books \\ that you can’t remember \\
    外婆做的火腿肠比外面\color{red}炸的酥 \\
    Grandma's ham sausage is crispier than \\ fried outside \\
    油烟的香味弥漫不那么\color{red}大的屋 \\
    The smell of lampblack pervades the not so  \\ big house \\
    外婆的故事总会让大人笑\color{red}着哭 \\
    Grandma's stories always make adults laugh \\ and cry \\
\end{tabularx}
\end{table} 

\paragraph{Translation of Chinese in Figure~3}~
\begin{table}[H]
\begin{tabularx}{\columnwidth}{l}
    我抬头仰望。天空的苍茫。\\
    I looked up. The sky is vast. \\
\end{tabularx}
\end{table}    

\paragraph{Translation of Chinese in Figure~4}~
\begin{table}[H]
\begin{tabularx}{\columnwidth}{l}
    是这座城市\color{red}{的气象}\\
    It is the weather of this city \\
    让你感受生命\color{red}{的力量} \\
    makes you feel the power of living \\
\end{tabularx}
\end{table} 

\paragraph{Translation of Chinese in Figure~6}~
\begin{table}[H]
\begin{tabularx}{\columnwidth}{l}
我长大的地方像一个简朴\color{red}的寨 \\
The place where I grew up is like a \\
simple village \\
公里也许大的远方简直是个小小\color{red}的寨 \\
A small far away village \\
偶尔穿件毛衣那样子\color{red}很可爱 \\
It is cute to wear a sweater occasionally \\
远方可单纯的姑娘还是单\color{red}纯的孩 \\
Is it a simple girl or a simple child far away \\
是放不下的故事大声的\color{red}喝彩 \\
Cheers loudly for the unforgettable story \\
像快乐的小孩莫名的敞\color{red}着怀 \\
Happy kids like happy kids \\
\end{tabularx}
\end{table}

\begin{table}[H]
\begin{tabularx}{\columnwidth}{l}
几百公里我们相约到未\color{red}来 \\
Through hundreds of kilometers,\\ we meet in the future \\
在那无尽的沙漠\color{red}和海 \\
In the endless desert and sea \\
看着温暖花\color{red}开 \\
Watching the warm flowers bloom \\
花一样\color{red}的在 \\
Like flowers be there \\
写动人的\color{red}天籁 \\
Write moving sounds of nature
跟着自由\color{red}自在 \\
Feeling the freedom \\
消沉\color{red}在那片海 \\
Sometimes depressed in the sea \\
不懂儿时的他们不懂\color{red}什么是爱 \\
I don’t understand their childish. \\I don't know what love is \\
到现\color{red}在你看来 \\
Till now you see \\
最\color{red}真的迷彩 \\
It is The most true fantasy \\
\end{tabularx}
\end{table} 
\end{CJK*}

\end{document}